# Zero-knowledge Based Proof-chain

A methodology for blockchain-partial system

Yuqi Bai
Department of Computer Science
Hebei Petroleum Vocational
Technical University
Chengde, Hebei Province, China
cdpc_byq@cdpc.edu.cn

Lei Luo
Department of Computer Science
Hebei Petroleum Vocational
Technical University
Chengde, Hebei Province, China
cdpc_ll@cdpc.edu.cn

## ABSTRACT

Intuitively there is drastic distinction between the "pure" decentralized block-chain systems like Defis and those that only utilizes block-chain as an enhancing technology but remains centralized with real-world business model and conventional technologies like database, application server etc. Our study explores extensively this distinction from a methodological point of view, classifies them into blockchain-complete and blockchain-partial, analyzes key features of the two types, and reveal the root cause of this distinction. We analyze the function or, in more strong words, the "ultimate purpose" of blockchain in the blockchain-partial systems, and present a conceptual model we named proof-chain that quite satisfactorily represented the general paradigm of blockchain in blockchain-partial systems. A universal tension between strength of proof-chain and privacy is then revealed and the zero-knowledge based proof-chain takes shape. Several case studies demonstrate the explaining power of our proof-chain methodology. We then apply proof-chain methodology to the analysis of the ecosystem of a collaborating group of blockchain-partial systems, representing the paradigm of public and private data domain whose border the proof-chain crosses. Finally, some derived guidelines from this methodology speaks usefulness of our methodology.

## KEYWORDS

Blockchain, methodology, proof-chain, zero-knowledge proof, privacy-preserving

## 1 Introduction

Blockchain is nowadays a far-reaching technology quite beyond its role played in crypto world and start having huge impact on extant traditional real-world systems with the list of its applications ever expanding[1]. Path of this expansion leads from cryptocurrency to more complex crypto financial organizations, then to meet and merge with the whole world of information systems that serving and supporting every aspect of our business and life, in private or public sector. It is not necessary to enumerate some or all of them, the point of significance here is that there is a milestone in this path of expansion worthy of great attention.

When we use decentralized ledger (blockchain) technology in information systems that support some business process like supply-chain, social credit-rating or any other systems that are the result of decades of endeavors of informatization in every aspect of our life, we are in a position which differs drastically (also essentially) with those in the new-born Defi (decentralized finance)[2]. A Defi organization is completely closed on the ledger, meaning all its transactions, activities or events are handled by smart contracts and recorded by the ledger; it is operated inside a secure "sandbox". But in those information systems only part of the business domain can be transferred onto distributed ledger, or in other words, it is not closed on ledger. This is not a trivial difference and not one of only technological nature but of great methodological significance. Our work is to analyzes the problems raised by this situation and recommend a solution called "zero knowledge-based proof chain".

To make notations simple, we will call systems like Defi "blockchain-complete" and those information systems that is not closed on-chain "blockchain-partial" in following text.

This division between blockchain-partial and blockchain-complete is intuitive but its signification of methodology will come to the surface with the unfolding of our work.

[1] LINDA PAWCZUK · ROB MASSEY, J.H. *Deloitte's 2019 Global Blockchain Survey: Blockchain gets down to business*. Deloitte [report] 2019; Available from: https://www2.deloitte.com/content/dam/Deloitte/se/Documents/risk/DI_2019-global-blockchain-survey.pdf.

[2] This distinction has already caused some attention and expressed quite clearly in a report from Deloitte
William Bible, J.R., Peter Taylor. *Blockchain Technology and Its Potential Impact on the Audit and Assurance Profession*. Available from: https://www2.deloitte.com/content/dam/Deloitte/us/Documents/audit/us-audit-blockchain-technology-and-its-potential-impact-on-the-audit-and-assurance-profession.pdf.

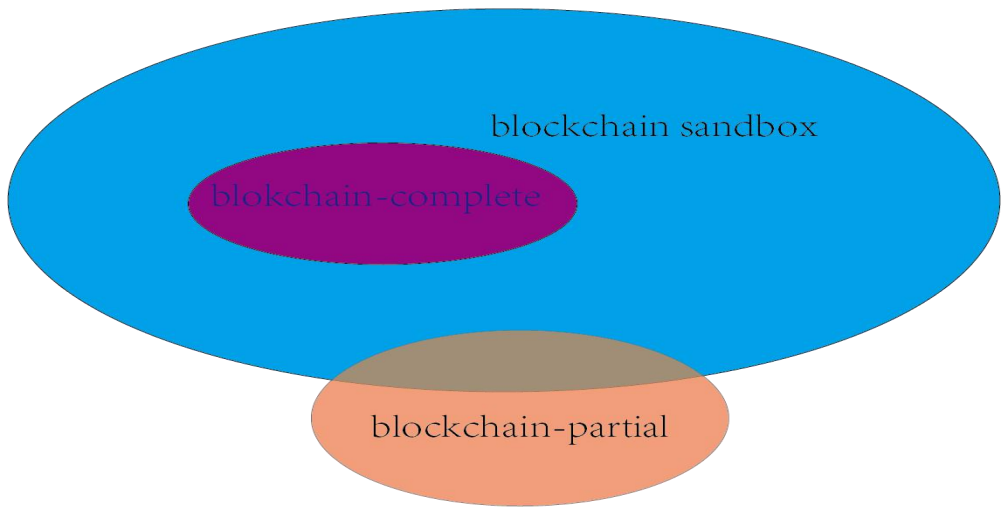

**Figure 1:** blockchain**-complete and blockchain-partial**

## 1.1 Blockchain-partial system is of relevancy

In methodologies of the so-called enterprise information system analysis and design[1], the key concept is something called "business model"[3] which is certain conceptual abstraction of the corresponding real world business entities, activities, events and processes; this model is a common language in the information system construction and evolution life cycle. From view-points of this conceptual business model, in blockchain-complete system, the whole implementation of the business model exists inside the secure blockchain, this is what we mean by saying it is closed on-chain or blockchain complete. Instead, for block-partial systems, only part of the business model extends into this secure sandbox, the rest still exposed to off-chain weakness. We use abstract business model as a measure to cut the line between the two type of systems because in concrete domain the boundary tends to be obscured and difficult to divide them: even systems with its business model completely inside the sandbox has various components off chain like user-interaction or some other peripheral parts, and from an abstract view they are supportive instead of core components and can be abstracted away, or, in languages that is more sophistical they are more being of representation character that is some representation of the essence expressed by the business model. This methodology makes things much clearer though we still cannot give technically precise delineation of the blockchain-partial and blockchain-complete systems. Defi organizations (MakerDao, Compound) are typical blockchain-complete systems because all its core transactions are on-chain, there is no off-chain business process to handle its functions like borrowing, lending or mortgage as n real world financial intermediaries, their business model is closed on-chain, they are blockchain-complete. A supply-chain system that utilizes blockchain as a ledger to store hashes for some key data is blockchain-partial for the main body of its transactions and data still exists in relational databases or some cloud storage. The boundary is quite clear.

It is obvious that when we step out of the utopia like Defi and try to utilize blockchain technology in some traditional information system supporting real world business, in private or in public sector, almost always we are handling blockchain-partial systems. The reasonableness of this pattern is evident because security by means of decentralization is not without cost, in fact it is extremely costive. (Some effort in the decentralization sector has been made to transfer whole application on-chain, the opportunity for its success is dubious and still remain to see [2]). Blockchain-partial systems takes half or more of the blockchain application domain. A methodology for blockchain-partial systems is vital for adoption of blockchain technology.

## 1.2 Previous research

Distinction between application of blockchain technology in defi systems and business process(the so-called enterprise application) is recognized and some particular issues like tension between immutability and upgradability is studied[3] [4] [5]. Blockchain in IOT[6] [7], supply-chain system[3, 8, 9], human resource management[10], medical-care[8, 11], are focuses of extensive research from alleged methodological perspective. The fundamental issue of the benefit or value of blockchain technology in certain scenarios also is touched topic[12].

We find something missed by these researches:

Where does the border of separation located that is of significance from methodological perspective? IOT, supply-chain system, human resource management and medical-care, all these specific systems raise specific issues when meeting blockchain; In fact, as we said before and analyzed in the following, the most important fact that is of most significance is when core activities, events or data of the system breaks out of the enclosed sandbox of blockchain, or in other words, when only part of the activities, events or data of the system is transferred onto blockchain. This distinction is what calls for a methodology with which particulars of systems in specific domain is better framed.

Enterprise applications as a far-reaching domain enclosing many expertized types of system but still is narrow from highest-level methodological perspective. SaaS platform, government-authorized social-credit rating systems, in one word, more large-scale systems, ecosystems in industry-scale or social-scale faces common issues because of its blockchain-partial property. More important is the fact that benefits or value of blockchain appears and becomes significant just when systems collaborate across borders of enterprise.

The division of permissionless and permissioned blockchain is an important aspect of blockchain-supported system[12] but not of major concern in our work.

## 1.3 Our contributions

1. It is time for Blockchain to become widespread choice of technology supporting traditional real-world information system whatever they are. Our work provides a general unified methodology for blockchain-partial systems enclosing all kinks of systems that is outside of the scope of the new-born pure on-chain crypto world, presently most of them defi organizations constructed by smart contracts. This methodology abstracts away from specific application domains.
2. The proof-chain concept catches crucial issues of blockchain-partial system and thus constitutes the most important insights of our work. Even the value or motive of using blockchain in blockchain-partial system is not always plain and clear for stake-holders, in what mechanism the blockchain has made or will make a blockchain-partial system trustworthy causes lots of confusion in users and designers during every phase of system lifecycle. The purpose and usage of proof-chain methodology is to make people clear-sighted; this is what a methodology means and why it matters.

[3] This word is used in context of system development instead of some commercial one.

3. Intrinsic conflicts between strength of proof-chain and privacy-preserving is revealed and a solution based on zero-knowledge is then presented.
4. A paradigm of private and public data domain is a highest-level view of block-chain partial ecosystem which demonstrating explaining power of proof-chain methodology.
5. Some practical guidelines are listed as derivations from our methodology.

# 3 Proof chain

## 3.1 Value provided by blockchain for blockchain-partial system

It is not necessary for us to dive into technical details to ask the question: Why we need blockchain as data storage in certain system? And further, why we are confident that the usage of blockchain in a system makes it trust-worthy? These are not questions that can be readily answered by just listing appreciable features of blockchain technology, decentralization, tamper-resistance, etc., just as a civil engineer cant said his work is worthy just because much steel is used. He is to explain from a structural mechanistic point of view instead of that of construction material.

For blockchain-complete systems like Defi, value provided by this technology is not quite problematic; they are executed inside this "secure sandbox" without anything crossing the border; there is no security hole and the protective intensity is uniform everywhere. But not for blockchain-partial system: part or main part of it has to remain outside of the protection of the blockchain as mentioned before. It is this situation that make value in blockchain-partial system problematic because firmness depends on the weakest link. Here is where we need a methodology without which the value of blockchain in blockchain-partial systems is clouded and dubious.

We cannot make people believe the building is worthy to live in just because we put some steel inside the wall, what is needed is some construction mechanics concerning usage of steel material. Blockchain is this steel material for information system design and construction, it is new material, its application will influence and change old construction theory, without this new theory the new material is useless.

In business practice, lack of methodology for blockchain-partial systems has already brought up quite many pseudo-blockchain applications and caused much waste of resources. A platform selling fake goods can put its data on blockchain but this will not make its fake goods more worthy.

In this section we view block-chain-partial and blockchain-complete system from perspective of conceptual business model and in this view the delineation is much clearer so we give the definition and a remark:

**Definition of Blockchain-complete and blockchain-partial**: If all data (transactions, activities, and events) defining the core business model of a system is on-chain, this system is block-chain-complete; otherwise, it is blockchain-partial.

**Remark 1**: Blockchain is construction material for information system as steel is material for buildings; we need a theory and methodology for using this new material as we need mechanics for building using steel.

## 3.2 Intrinsic natural order in transactions, events, activities and data, and the concept of proof chain

In blockchain-partial systems only part of its business model is on-chain, what makes this business model is transactions, events, activities, and data, it is a set of transactions, events, activities and data, and what blockchain-partial system does is to transfer some of them onto the blockchain. The question is how to select, the general principle of this section is what we call methodology of blockchain-partial system. To be less verbose we will use "data" to notate transactions, events, or activities.

Here the key point is: data is not some chaotic entities; they have intrinsic natural order between them. meaning of this order might be of time-dependent physical cause-effect or logical or highly relevant from statistical view. Physical cause-effect and logical relationship is naturally directional, statistical relevant relationship can be made directional in context, so we can view all of them as directional, and whatever properties this relationship between data entities is of, existence of starting entities constitutes proof of existence of the end, or, to put it more precisely, the integrity of one constitutes proof of integrity of the other. This proof linkage between data entities further makes proof chain.

## 3.3 Chronology and detail-summary in proof-chain

Although we cannot enumerate all possible features of the order or linkage of the proof-chain, it almost necessarily implies some chronological or detail-summary meaning, and entities that recorded earlier always be the one that is more detailed, or vice versa. the proof-chain will present its value. put it simply: Attacker have much more difficulty to compromise the system if he has to put his hands on the detailed data at a much earlier time point to prepare for a future attack. Just imagine someone that is required to provide five years tax capability to be qualified for something. He has to be a prophet to plan for this years ago and this further will be much harder due to the task to handle many detailed data entities. Back upon the proof chain difficulty level for attacker rises precipitately and when this difficulty reaches some limit it becomes economically infeasible for attackers to do it.

It is this mechanism of proof-chain that constituents the fundamentals of the blockchain-partial system.

## 3.4 Some insights

Proof-chain makes it clear that application of blockchain technology is not a black-white issue; instead, it is an issue of broad spectral with blockchain-complete systems on one extreme and applications with none or almost meaningless on-chain proof on the other extreme. Between them value of blockchain proof should be considered and analyzed in light of strength of the proof chain which is in positive relationship with its length. Longer length of proof chain implies more security, which level of security is necessary is contingent on specific domain problems. Proof-chain is a methodology, not a solution for blockchain application, solutions are principled by it.

We say that data entities are not a chaotic but is ordered by intrinsic natural order, in fact this order is what gives the data topological structure and constitutes an integrated "business model", and proof-chain is a sub-graph of this topology or business model. In light of this, proof-chain of Blockchain-complete system is this full topology or the business model itself.

It is not inappropriate to say that blockchain-complete system is so called because the complete business model is closed on-chain, of course not in a causal sense but just a synonym of the same concept. Whether a system is blockchain complete or partial is due to some essence or intrinsic nature of it. Defi can be and is closed on-chain, the root cause, or to be more precise, the rationality of this is that Defi system is the combination of the effort of financial freedom and technological potential (smart contracts) created by blockchain. Defi is long-time aspired economic Utopia made possible by technology. But the scenario is totally different for blockchain-partial system, here blockchain is taken as a new technological element for existed or traditional information system supporting real world business: supply-chain, auditing, credit-rating etc. The business of this kind of system is part of real world, from a metaphysical viewpoint there is no clear-cut border separating them form real world. But for such economic utopia like Defi the border do exists. This difference of the two types means a lot for their proof-chain: Proof-chain in blockchain-complete systems terminate at edge of the utopia, but in blockchain-partial systems it never really terminates in metaphysical sense, but is terminated when enough security is obtained.

**Remark 2**: proof-chain is a sub-graph of the topology of business entities or, put it simply, sub-graph of business model

**Remark 3**: Whether a system is blockchain complete or partial is due to some essence or intrinsic nature of it

# 4. proof -chain

## 4.1Conflict between privacy-preserving and proof-strength in proof chain

When proof -chain grow to its upper stream, chain length become longer, proving power increments, attacks become more difficult even infeasible. This scenario is what we described until now, here a constraint emerges. As we said before, data on upper position of proof-chain generally tends to be more detailed and detailed data tends to be on the verge of invading privacy, personal or commercial. Strength of proof-chain conflicts directly with privacy-preserving.

This conflict is inherent and ubiquitous. In application domains like auditing and credit-rating detailed financial data on upper stream of proof-chain is heavily private. Citizens or organizations have responsibility to expose statistical report of their financial conditions to be credited in the mean time keeping their everyday bank or tax record secret. In supply-chain transparency theory and practice parties (manufacturers, logistics, wholesalers, and retailers) are encouraged to publish their data, but here also a balance between transparency and privacy exists. The inherent and ubiquity of this conflict is a fundamental aspect of proof-chain

## 4. 2 Zero-knowledge proof as a solution for the conflict

zero-knowledge proof is the choice everywhere privacy-preserving is critical, so for our proof-chain. This article is not about ZKP technology itself but to study it from an methodological viewpoint, its core functionality in solving the intrinsic conflicts of the proof-chain as core methodology for blockchain-partial systems, no time is wasted in any introductory of it because enough material can be found for readers[4][5]. What we will do in following is to consider some typical application scenarios to frame the general mechanism of zk-based proof-chain.

## 4.3 Case studies of zero-knowledge based proof chain

**Identity Information Service**

As we said before, the root cause that makes a system blockchain-partial is that it is modelling part of real-world business, and this fact also makes anonymity almost impossible in most application case, thus identity is one of its critical elements and is actively researched[13, 14]. Every blockchain-partial system takes "Identity wallet" as its functionality entry point. Users will have to provide the necessary part of their personal information (name, age, sex, location etc.) to be qualified for some social or commercial goods, like in requesting certification or licenses for some investment instruments or to be credited for something. This raises the question:

1. To prove that the information comes from authoritative sources.
2. Only necessary information instead of sophisticated personal archive is presented.

Here the structure of the proof-chain is "personal archive to personal information digest", the authority of some national archive manage center terminates the chain and no further back tracing to some historical story. In such system some well-known authority with good traditional common social confidence provides practical and acceptable limit of proof chain This is what we have analyzed previously.

[4] Goldreich, O. & Oren, Y. Definitions and properties of zero-knowledge proof systems. *Journal of Cryptology* **7**, 1-32 (1994).

[5] Goldreich, O. & Krawczyk, H. On the composition of zero-knowledge proof systems. *SIAM Journal on Computing* **25**, 169-192 (1996).

The zero-knowledge proof solution of this proof-chain is described:

1. Authoritative archive manager put hash of someone's personal archive on chain (like some smart contract invocation) and will update it when necessary.
2. Some service provider (maybe) needs to know age of its customer and the service provider request the service API of the authoritative archive manager to get age with zero knowledge proof
3. The service provider does verification with the proof (including age) by invocation of smart contract. The smart contract does it by the proof and archive hash that is managed by the Authoritative archive manager.

We implemented the process described above using zk-Snarks but here technical steps like "setup" phase is not mentioned. Technical details of zero-knowledge implementation is not essentials of our topic.

This might be viewed as the simplest form of zero-knowledge based proof-chain, it depends directly on some authority as its last resort. People tends to consider everything related with blockchain as being certain kind of pure decentralization with no authority stepping into this utopia, but at least this is not the case for blockchain-partial systems, authorities are good and necessary anchor point for system and in our topic, termination point for proof chain. The specific meaning of "authority" itself is contextual, it might be administrative institutions with little or huge or any extent of arbitrary power or something based on representative democracy or some professional societies with good tradition whatever. It is not reasonable for us to take values originated from decentralization of blockchain as the only one worthy and throw away existed values away. In blockchain partials, values of blockchain are more of a frictionless transferring carrier of trust instead of origin of trust.

**Auditing**

To some extent, auditing is itself a process of building up a proof-chain and the proof -chain mechanism or methodology can be viewed as methodological generalization of auditing in domains other than financial auditing. Thus Zero-knowledge comes to be the suitable mathematical tools to solve the conflict of proof-strength and privacy-preserving in auditing[15-17].

[17] constructs a stereotyped accounting system of banks that leverages on-chain mechanism and off-chain databases and zero-knowledge； what makes it conspicuous from methodological view is its effort to separate part of the business of real banks and makes a blockchain-complete system. Still the root cause as we analyzed previously does not vanish by this effort and such a stereotyped system does not fit into realistic financial auditing landscape; auditing supported by blockchain is essentially blockchain-partial as described by the report from Deloitte[6]. Accounting is not a profession suitable for some specialized distributed ledger to contain every auditing-related data of any real-world business of real bank which is different with Defi system. What we do is to find some suitable proof-chain;

In fact, from view point of proof-chain, the zero-knowledge auditing data and computations in[17] consist a proof-chain, only if we open up its closedness like checking integrity of all transaction (This kind of closedness exists only in a narrow-minded paly-around one: all banks and all business of all banks is transferred on-chain overnight; with one piece left off-chain the closedness is lost). This auditing proof-chain stops at everyday detailed record and Description:

1. Hashes of everyday accounting data go on-chain.
2. Auditor requests for statistical report from audited entity and get result with proof.
3. Auditor invocate smart contract to verify this result.

Specific technological respect worth mention. Granularity of data blocks of everyday records is a design problem; some intermediary statistical report might be made on-chain(hash) to save time and resources; zk-Snarks is an alternative to those in [17].

The structure of proof-chain in auditing is "everyday records to statistical reports" whose strength lies in the time-span and the detail both making it much more difficult for attackers to plan and implements an effective attack. This is in stark comparison with those in the identity information service case which take some confidence in authority as termination point for proof-chain.

**Supply-chain transparency**

Supply chain is typical in its crossing-organization feature for which the problems of trust arise to considerate level which makes blockchain quite attractive; transparency of supply chain[18-20] means a lot: Trackability of goods, credit-rating of entities are all aspects under the hood of transparency. As is said in [19], "Blockchain cannot supersede the existing technologies, but can complement them", the transparent supply chain necessarily be blockchain partial and proof-chain be the method in analyzing it. Nothing particular for this.

What is interesting is, transparency, as a targeted value of supply chain, will push the border of privacy of supply chain entities (manufacturer or producer, logistics, suppliers and retailers) to more contracted area, but of course also balanced between them:

1. More transparency means less privacy, a balance is reached. Data is divided into transparent domain and privacy domain.
2. Enough strength of proof-chain is necessary which is not guaranteed to be with transparency domain. the objective in

[6] William Bible, J.R., Peter Taylor. *Blockchain Technology and Its Potential Impact on the Audit and Assurance Profession*. Available from: https://www2.deloitte.com/content/dam/Deloitte/us/Documents/audit/us-audit-blockchain-technology-and-its-potential-impact-on-the-audit-and-assurance-profession.pdf.

design of Proof -chain and those of transparency in independent of each other.
3. If proof chain stays within limit of transparency, the privacy domain is not invaded and no zero-knowledge proof needed.
4. If proof-chain extends into the privacy domain, data within this domain has a responsibility to prove data within transparent domain.

This relationship between transparency, privacy and proof-chain is illustrated in figure 3. This analysis is a demonstration of the explaining capability of our proof-chain as a methodology, and for methodology the power of interpretation means its power in directing and guiding system constructing activity.

Some problems concerning proof-chain in supply chain

In most cases, what the supply chain entities exposed is statistical report of operational record data of the system, or data in transparency domain tends to be statistical report (data in privacy domain tends to be operational transaction records). The meaning that the transactional data is proof of statistical report is the same with those in auditing. What the proof-chain in supply-chain transparency can do and actually do is to guaranteed that those shared data originated from real transactions, detailed operational transaction is the anchor point of proof-chain, no further anchor points is available.

A corrupted supply chain ecosystem with all its members profiting from an economy of fake goods will not be restrained only by such proof-chain. This is another conclusion made clear by our methodology which is not less important in significance due to the danger of blockchain being utilized as a pretense.

**Privacy-preserving Prediction model-weak proof link**

Semantics of the proof chain might be heterogenous: causality or logical necessity(mathematics), or of a Bayesian probabilistic nature, in the adoption of certain prediction model. Machine learning, deep or not deep, makes up another constructive mechanism for proof-chain that needs some deliberation from these methodological viewpoints.

Enough strength of chain link is preamble and of course, contextual to application scenarios. Interpretability of this prediction model is critical without which input features of model cannot be treated as proof of output of this model, like those models that compute credit-ratings for social entities. Ubiquitous confidence or acceptance of this model should have been established. Every effort of enhancement made, this proof link being set up by correlation is essentially weaker than physical causality or logical and mathematical reference.

The zero-knowledge problem for proof-chain in ML prediction model is a sophisticated domain under intense research. [21] “develop an end-to-end optimizer that compiles a floating-point pyTorch model to R1CS constraints”, and the R1CS constraints then can be translated to quadratic arithmetic program (QAP) to fit into zk-Snark environments like libsnark or zokrates.

## 4.4 Pattern of zero-knowledge link

We identify a pattern of zero-knowledge link in proof-chain from our case studies deserving some description and interpretation. We say “link” since here the focus is on one piece of the proof-chain, a connection from preserved data to shared ones. The pattern consists of four components: preserved data, shared data, blockchain and zero-knowledge enabled computation unit.

1. preserved data
2. shared data
3. blockchain
4. zero-knowledge enabled computation unit

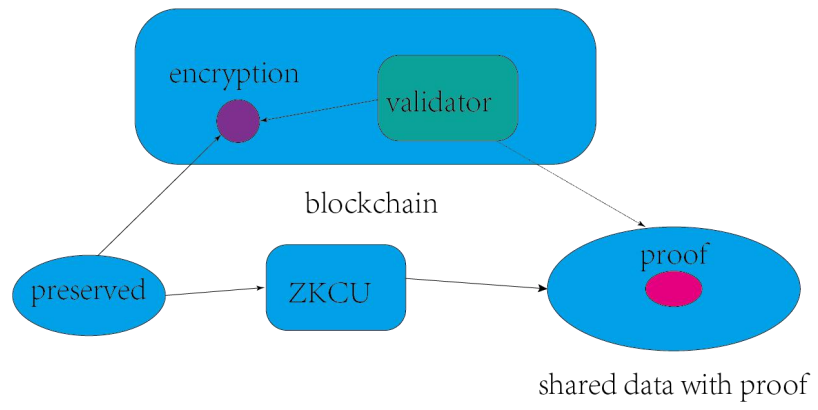

**Figure 2:** pattern of zero-knowledge link

function of this pattern:

1. the preserved data is encrypted and put onto blockchain at appropriate time point which is an issue of design
2. zero-knowledge enabled computation unit takes preserved data and output the shared data with proof
3. a validator takes preserved data encryption and the shared data with proof to validate that the shared data is a trusted result of right data source and right computation.

This is a rough outline of a pattern in need of quite a lot of explanations, in fact it is a conceptual framework, an abstract pattern instead of a concrete one in design activity.

1. The preserved data tends to be transactional and is created and maintained by conventional off-chain business system; they are in a sense “first-hand” which makes them certified as proof to second hand ones. The opportunity of their being encrypted and put onto blockchain is design decision of the business transactional system. Another issue is the granularity of the data encryption which has big impact on computation and validation logic and is vital for design but is not of concern for our methodology.
2. The ZKCU (zero-knowledge enabled computation unit) is much more abstract component of the pattern; it might be as simple as taking one field from a data structure or complicated as execution of some prediction model. Type of zero-knowledge technology used also varies.
3. The validator illustrated in figure 2 is implemented by smart contract but an off-chain implementation is acceptable and reasonable in some scenario, the choice also an issue of system design.
4. Shared data and its proof might be persisted with some updating arrangement preventing high-frequency invocation of ZKCU. Another consideration for persistence of shared data is its participation in further computation to extend the proof-chain.

## 5 Paradigm of zero-knowledge based proof-chain ecosystems

The problem of trust arises when data is shared between organizations and blockchain is alleged to be the technology in solving this cross-organization trust problem. Cross-Organization collaboration or data-sharing is the target for blockchain without which this technology lost its proper and becomes meaningless and futile. Blockchain provide a trust infrastructure above which a cross-organization ecosystem thrives. This is quite obviously demonstrated by our supply-chain transparency case study for in it the collaborations and data-sharing between supply-chain entities(members) is made explicit, but in all other domains where blockchain played its appropriate part, this paradigm emerges repeatedly. The division of data into transparent domain and privacy domain and the existence of zero-knowledge based proof-chain with this context is a generic paradigm. Let us change some word to make things looks more general: we use public and private instead of transparency and privacy to abstract away from the supply chain transparency context:

1. Private data domain: data in this domain tends to be internal, operational, more detailed with finer granularity, and transactional, they are generated locally by a system and tends to be privacy-sensitive. They also prone to strict constraint by business rule and this constraint makes them much more difficult to be modified with integrity undamaged.
2. Public data domain: data in this domain are shared or exposed to collaborators or thoroughly public and tends to be some digest or statistical computation of the data from private domain.
3. Proof chain links the public domain to private domain, demonstrates that the link is a correct one. Strength or semantics of this link are contextual, being logical, mathematical, causal, or correlational, securitized by system stake-holders to evaluate if it is strong enough for the specific business supported.
4. Zero-knowledge proof makes the border between private and public unchanged in implementing the proof link.
5. From viewpoint of trust, proof-chain makes attacks much more difficult by transforming attacks to data in public domain into attacks to private domain where there are much more authority or business-rule constraint and with a much longer time-span.
6. The public data domain exits in the service API layer of the member of ecosystems and they constitute the infrastructure of this ecosystem, block-chain is elements of this infrastructure and functions through zero-knowledge based proof-chain. This is the highest-level big picture of the paradigm.

All the cases we studied above fits into this general paradigm.

## 6 Pragmatic guidelines and Principles

If a system is Blockchain-partial for good reason, efforts to make it blockchain-complete are almost always futile, the right thing is to accept this imperfect reality and find ways to compromise with it. This division of partial and complete is not superfluous but reveals essentials.

In situations where no cross-organization data-sharing exists the necessity of blockchain is dubious. We usually do not give proofs in administrative activities to people within a hierarchical ordered organization. Only organization-crossing makes the problem of trust conspicuous enough for trust enhancement by technology like blockchain indispensable.

If a blockchain-partial system does not present a proof-chain that is sufficiently strong, then the worthiness of blockchain in it tends to be dubious; if the proof-chain is open to simple fraudulence, this is a silly blockchain project no matter how heavily blockchain is utilized. A proof-chain mental testing and discussion is critical before and during lifecycle of system. If no hope of finding valid proof-chain for a business it means blockchain is useless here, at least before you can find that proof-chain.

The utilization of Zero-knowledge proof technology should be scrutinized under the proof-chain methodology. We find a project that use zk-Snarks to prove to supervisors that some ship has not crossed over border of some geographical area but the input of location is almost free to fake, no proof-chain design to make this kind of fraudulence less easy.

Finding established authority is important for design of proof-chain because it is good anchor points of trust and termination point of proof-chain. This is where blockchain compromise with authority in real-world.

Semantics and strength of proof-chain are both contextual and should be paid enough attention in design. Correlation is incorporated into proof-chain when prediction model from machine leaning or deep learning is adopted and this is of great significance and acceptance of the correlation as proof is not a straight-headed question. Good interpretability and massive acceptance are crucial.

Last, Zero-knowledge proof is not necessarily needed in specific proof-chain when privacy is not so sensitive although it is in most cases. Save it if situation permits because zero-knowledge is expensive in cost and limited in ability of business modelling, still.

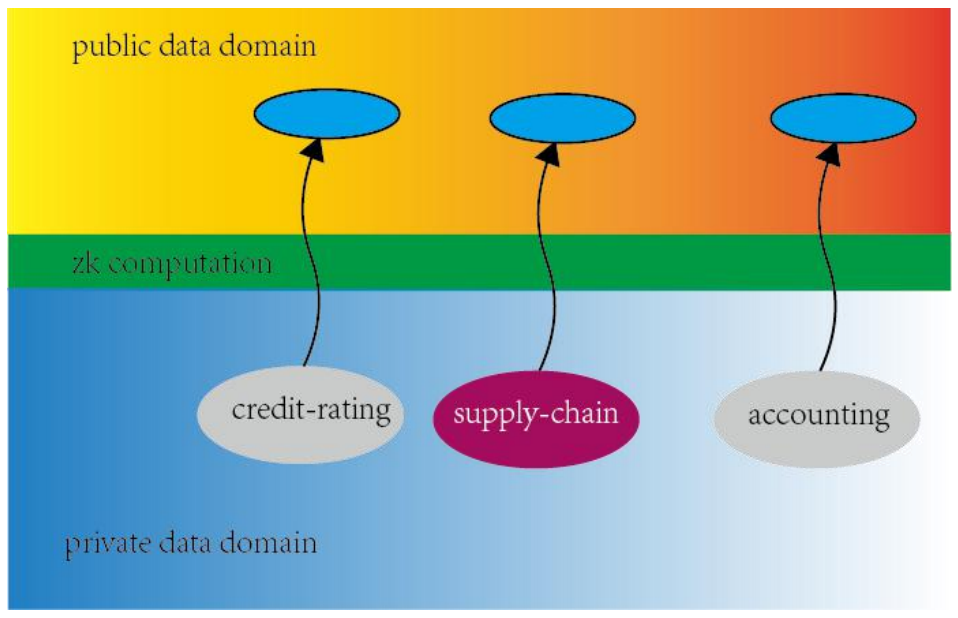

**Figure 2:** pattern of zero-knowledge link

**Remark 4**: The big story behind proof-chain is organization collaboration and data-sharing in an ecosystem with proof-chain making the shared public data trusted by its root into operational private data

## 7 Limit of applicability of our methodology

Blockchain technology and its application is in the process of rapid expanding and deepening, it is an evolving landscape. The co-existence of blockchain-complete and blockchain-partial is in a pattern of fluid and morphism.

More radical practitioners tend to disrupt existent business model and start from scratch[7], and this tends to be the promotion of business model complete on-chain because of the freedom from constraint of the legacy. This is one force that pull and push blockchain-partials to blockchain-completes.

We analyzed the root cause behind blockchain-partials in previous section to demonstrate that blockchain-partial or not is not an issue of complete arbitration, this is another force speaking for blockchain-partials.

From pure view pint of rationality regardless of legacy, some business model as a whole naturally legitimizes blended architecture of blockchain and conventional technology, this fact is the necessary precondition that blockchain-partial exist, but not one of sufficiency. If those parts of business model that is suitable to be on-chain is well-defined into some self-sufficient box, the proof-chain breaks and there emerges blockchain-complete subsystems embedded in the container business model. This embedding architecture is much preferred and is one force from blockchain-partials to blockchain-completes, but whose realization is also a phenomenon of ecosystem evolution.

Blockchain -partials tends to evolve into a mixture of clear-cut divisions of an off-chain business model and embedded blockchain-complete subsystems, but this is not the end of the story. It will always be necessary for the off-chain business model to share some of its data just as demonstrated by the private and public data domain paradigm of ecosystem, so blockchain-partials do not perish and our proof-chain methodology keeps its territory of applicability.

Blockchain-partial systems is also a phenomenon of ecosystem evolution. Our methodology is applicable whenever and where ever it appears in this ecosystem.

[7] LINDA PAWCZUK， ROB MASSEY, J. H. Deloitte's 2019 Global Blockchain Survey: Blockchain gets down to business. Deloitte (2019).
"Where enterprise organizations seek ways to integrate blockchain into their existing business models—or, more accurately, how to transform existing processes and systems to work with blockchain—emerging disruptors built their businesses around blockchain from the start"

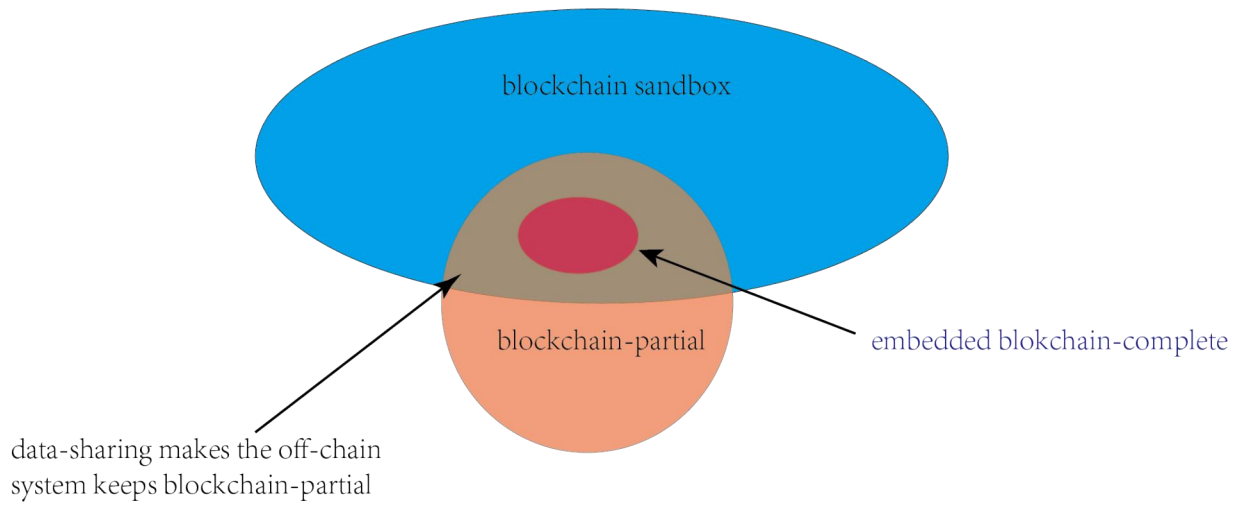

**Figure 2:** embedded blockchain-complete and blockchain-partial

**Remark 5**: blockchain-partial systems and proof-chain methodology survive with evolution of the ecosystem

## Confidential computing or Privacy computing

We are not ambitious enough to allege that proof-chain is a suitable methodology for those forefront domains like Confidential computing and Privacy computing[22-27], although the paradigm in these applications can be fixed into the zero-knowledge based proof-chain pattern just by take the computation as a kind of proving process. But here in these next-generation data-oriented applications the is no issue as raised by integration with legacy business model and value of our methodology in this context is questionable.

## Conclusion

Proof-chain as a methodology for blockchain-partial system demonstrates its explanatory power; it reveals essentials lurking deeply inside the entangle of confusing issues when blockchain technology is taken to be integrated into a complex of business model. This lucidness from methodological view of blockchain-partial systems is not a solution itself but helps greatly in searching the right solution in specific domain.

Our research of proof-chain methodology is extensive, reaching out to every important aspect:

1. From perspective of conceptual business model, we get the insight that essentially proof-chain is a subgraph of the whole business model that is t be enhanced by blockchain.
2. The semantics and strength of proof-chain is analyzed
3. Intrinsic tension of proof-strength and privacy is revealed and a zero-knowledge based solution
4. The private and public data domain paradigm of proof-chain from the level of ecosystem provides clear-sighted view of this ecosystem and is of significance
5. A set of pragmatic guidelines is derived which is helpful for practitioners.
6. Its condition of applicability is studied in context of ecosystem evolution.

## Future work

Our team is participating several blockchain-enabled projects of local governments which engages in building up platforms for social credit-rating. This application of blockchain is typical of blockchain-partial and suitable public-sector use-cases of our proof-chain methodology. We adopt libsnark and zokrates as implementation of the zero-knowledge part for our proof-chain.

Blockchain technology and its application is of high priority in China propelled by government policy; more important, the application of this technology in refactoring current business is greatly encouraged, in dark contrast with a policy of banning cryptocurrency. This is where blockchain-partial systems take place. But there is situations that is quite worrisome, even dangerous： the lack of a general theory, a methodology and guidelines for this thriving movement. Our work is meant for this.

## ACKNOWLEDGMENTS

We are grateful for the support from university administration and computer science department; the cooperation of local government in applying our methodology in the public-sector blockchain-enabled projects.